\documentclass[prd,twocolumn,superscriptaddress,nofootinbib]{revtex4-2}
\usepackage{natbib}
\usepackage{graphicx}% Include figure files
\usepackage{bm}% bold math
\usepackage{amssymb}
\usepackage{color}
\usepackage{amsmath}
\usepackage{mathrsfs} 
\usepackage{amstext}
\usepackage[english]{babel}
\usepackage{latexsym}
\usepackage{array}             % Tabular
\usepackage{multirow}         % Multicolumn Tables
\usepackage[usenames,dvipsnames]{xcolor}
\usepackage[colorlinks=true,citecolor=Blue,linkcolor=RubineRed,urlcolor=Blue]{hyperref}
\usepackage{subfigure}
\usepackage{mhchem}

\newcommand{\beq}{\begin{equation}}
\newcommand{\eeq}{\end{equation}}
\newcommand{\beqa}{\begin{eqnarray}}
\newcommand{\eeqa}{\end{eqnarray}}

\begin{document}

\title{Interface Dynamics of Strongly interacting Binary Superfluids}
\author{\textbf{Yu-Ping An}}%
 \email{anyuping@itp.ac.cn}
 \affiliation{CAS Key Laboratory of Theoretical Physics, Institute of Theoretical Physics,
Chinese Academy of Sciences, Beijing 100190, China}
\affiliation{School of Physical Sciences, University of Chinese Academy of Sciences, Beijing 100049, China}
 \author{\textbf{Li Li}}
\email{liliphy@itp.ac.cn}
 \affiliation{CAS Key Laboratory of Theoretical Physics, Institute of Theoretical Physics,
Chinese Academy of Sciences, Beijing 100190, China}
\affiliation{School of Physical Sciences, University of Chinese Academy of Sciences, Beijing 100049, China}
\affiliation{School of Fundamental Physics and Mathematical Sciences, Hangzhou Institute for Advanced Study, University of Chinese Academy of Sciences, Hangzhou 310024, China}
\affiliation{Peng Huanwu Collaborative Center for Research and Education, Beihang University, Beijing 100191, China}

 \author{\textbf{Chuan-Yin Xia}}
 \email{chuanyinxia@foxmail.com}
 \author{\textbf{Hua-Bi Zeng}}
 \email{hbzeng@yzu.edu.cn}
 \affiliation{Center for Theoretical Physics , Hainan University, Haikou 570228, China}
 \affiliation{  Center for Gravitation and Cosmology, College of Physical Science and Technology, Yangzhou University, Yangzhou 225009, China}

 \vspace{1cm}

\begin{abstract}
Understanding the interface dynamics in non-equilibrium quantum systems remains a challenge. We study the interface dynamics of strongly coupled immiscible binary superfluids by using holographic duality. The full nonlinear evolution of the binary superfluids with a relative velocity shows rich nonlinear patterns toward quantum turbulence, which is reminiscent of the quantum Kelvin-Helmholtz instability. The wave number of the fastest growing modes $k_0$ extracted from the interface pattern yields a non-monotonic dependence of the relative velocity, independent of the temperature and interaction. The value of $k_0$ first increases with the velocity difference and then decreases, which stands in sharp contrast to the results of mean-field theory described by the Gross-Pitaevskii equation and is confirmed by using the linear analyses on top of the stationary configuration. We uncover that the critical velocity associated with the maximum correspond to the case when the mean separation of vortices generated by interface instabilities becomes comparable to the vortex size, which could be a universal physical mechanism at strongly interacting superfluids and is directly
testable in laboratory experiments.

\end{abstract}

%\pacs{ 05.30.-d;73.43.-f;05.30.Pr}
\maketitle

\textbf{Introduction--}
Interface instabilities are ubiquitous and are of fundamental interest in fluid dynamics, biological systems and engineering applications. A well known one in classical fluid mechanics is Kelvin-Helmholtz instability (KHI) that occurs when there is relative velocity in a single continuous fluid or a velocity difference across the interface between two fluids.
This instability, in turn, produces waves, which typically leads to roll-up patterns in the nonlinear stage. Typical examples include the cloud formations on Earth and the Red Spot on Jupiter. There have been growing interests in KHI in quantum fluids. The quantum KHI could play crucial roles in many important  phenomena, ranging from the laboratory to astronomical scales, \emph{e.g.} the pulsar glitches
of rotating neutron stars and
the vortex formation in atomic Bose-Einstein condensates (BECs). Due to the quantum characteristics of quantum fluids (\emph{e.g.} vortex quantization), quantum KHI yields novel nonlinear dynamics that have not been well understood. An ideal testing ground for quantum KHI is superfluids characterized by the vanishing small viscosity. The first
experimental observation of quantum KHI
was made in~\cite{QKH3,QKH4}. It was found that in the presence of shear flow between the A and B phases of superfluid $\ce{^{3}He}$ in a rotating cryostat, vortices penetrate from the A phase into the B phase due to KHI. Another natural candidate to study the quantum KHI is the two components BEC which now can be produced in laboratory~\cite{experiment1,experiment2,experiment3,experiment4}.

Various studies related to the quantum KHI have 
appeared in the literature both experimentally and theoretically \cite{QKH1,QKH2,QKH3,QKH4,QKH5,QKH6,crossover,pattern}. Dynamical instabilities at the interface between two BECs moving relative to each other were investigated using effective theories in the absence of dissipation~\cite{crossover}. The wave number of the most unstable mode $k_0\sim v^2$ for small velocity
difference $v$, which is reminiscent to classical KHI, while $k_0\sim v$ for large $v$, where counter-flow instability dominates. After adding dissipation phenomenologically, Landau instability caused by excitation of negative energies occurs in addition~\cite{QKH2}. Besides, by controlling the  relative velocity and the coupling strength between the two components, various patterns can develop~\cite{pattern}. 

Nevertheless, most of these 
efforts rely on Gross-Pitaevskii equation (GPE) which is a model equation for the ground-state single-particle wavefunction in a weakly interacting BEC. In reality, intraspecies of the superfluids might be strongly coupled, and finite temperature effects and dissipation should also be accounted. 
Holography offers us a natural tool to include all those ingredients.  The strongly coupled quantum many-body systems at finite temperature and dissipation are encoded to gravitational systems of black holes with one higher dimension. Holographic superfluids have been widely studied in the literature, such as superfluid turbulence~\cite{Adams:2012pj,Du:2014lwa}, dark solitons~\cite{Guo:2018mip} and Kibble-Zurek mechanism~\cite{Chesler:2014gya,Sonner:2014tca,delCampo:2021rak}. The comparison between holographic superfluids and GPE~\cite{GP_compare1,Yan:2022jfc,Yang2023} were made as efforts to connect holographic predictions with experiments. Moreover, thus far there are few investigations on the interface dynamics for holographic superfluids. 

In this work, we study the interface dynamics of binary strongly coupled superfluids using holography. The stationary patterns obtained form our holographic model share some similarities with GPE. Nevertheless, we focus on the temperature effect on the interface dynamical instabilities far from equilibrium and uncover distinct behavior from GPE~\cite{crossover}. Our investigation unveils a remarkable phenomenon: the most unstable mode $k_0$ is non-monotonic as the velocity difference $v$ increases, irrespective of temperature. It first increases, arrivals at a maximum, and then decreases as $v$ is increased. Interestingly, we find that the peak location corresponds to the critical case when the average distance of quantized vortices generated along the interface is comparable to the characteristic vortex size. This not only provides a smoking gun for the difference between holography and GPE, but also uncovers novel underlying mechanism responsible for interface dynamics in strong coupling regime. We now discuss in more detail how we arrive at these results.

\textbf{Holographic model--}
We consider a $(3+1)$-dimension bulk theory that holographically describe the interface dynamics of two-component strongly coupled superfluids in two spatial dimensions. 
    \begin{equation}\label{action}
        \begin{split}
             \mathcal{L}=\frac{1}{2\kappa_N^2}(R&+\frac{6}{L^2})+\sum_{i=1}^{2}-(\mathcal{D}_\mu\Psi_i)^*
            \mathcal{D}^\mu\Psi_i-m_i^2|\Psi_1|^2\\&-\frac{\nu}{2}|\Psi_i|^2|\Psi_2|^2-\frac{1}{4}F^{\mu\nu}F_{\mu\nu}\,,
          \end{split}  
    \end{equation}
with $R$ the Ricci scalar, $L$ the AdS radius and $\mathcal{D}_\mu\Psi_i=(\nabla_\mu-i e_i A_\mu)\Psi_i$ ($i=1,2$). It involves two complex scalar field $\Psi_i$  charged under the $U(1)$ gauge field $A_\mu$ with its strength $F_{\mu\nu}$, see~\cite{Basu:2010fa,Cai:2013wma} for early studies. There is a direct interaction between the two scalar, mimicking the interaction between two components of superfluid. 

Working in the probe limit where the  backreaction of the matter fields is neglected, we fix the bulk geometry as the Schwarzschild AdS black hole
    \begin{equation}\label{backg}
        ds^2=\frac{L^2}{z^2}(-f(z)dt^2-2dtdz+dx^2+dy^2)\,,
    \end{equation}
where $f(z)=1-(z/z_h)^3$ with $z_h$ the location of the event horizon.  It corresponds to a heat bath with temperature $T=3/(4\pi z_h)$ on the boundary system. Without loss of generality, we set $L=1$ and adopt the radial gauge $A_z=0$. For simplicity, below we choose $m_1^2=m_2^2=-2$, $e_1=e_2=1$. Then, near the AdS boundary $z=0$, asymptotic expansions for matter fields read 
    \begin{equation}
        \begin{aligned}
            A_\mu=a_\mu+b_\mu z+\dots,\quad
            \Psi_i=\Psi_i^{(s)} z+\Psi_i^{(v)} z^2+\dots\,.
        \end{aligned}
    \end{equation}
From the holographic dictionary, we turn off the leading source term, \emph{i.e.} $\Psi_i^{(s)}=0$ and thus $\Psi_i^{(v)}$ corresponds to the superfluid condensate $\mathcal{O}_i $. Moreover, $a_t=\mu$ is the chemical potential and $\bm{a}=(a_x,a_y)$ are related to the superfluid velocity $\bm{v}_i\equiv(v_x, v_y)_i=\nabla\theta_i-\bm{a}$, where $\theta_i$ is the phase of the condensation $\mathcal{O}_i$ We have used bold-face letters to denote vectors in boundary spatial directions. In practice, we choose $a_x=a_y=0$, such that the superfluid velocity is given by $\bm{v}_i=\nabla\theta_i$ for each component. Throughout the paper we will keep the chemical potential of the system fixed. The system enters a superfluid phase below some critical temperature $T_c$ when the order parameter spontaneously develops a nonzero expectation value, which in the gravitational description corresponds to the scalarization of $\Psi$.

\begin{figure*}[htpb]
        \centering
        \vspace{-0.35cm}
        \subfigtopskip=2pt
        \subfigbottomskip=2pt
        \subfigcapskip=-5pt
        \subfigure[Normalized $|\mathcal{O}_1|^2$ for different $\mu$ with $\nu=1$ and $v_y=0$.]{
            \includegraphics[width=0.3\linewidth]{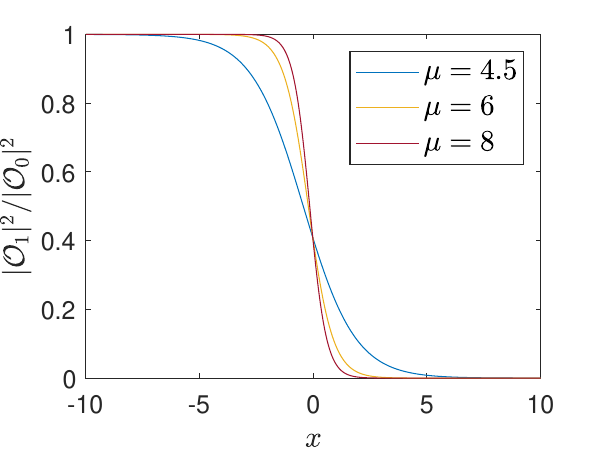}}
        \subfigure[Normalized $|\mathcal{O}_1|^2$ for different $\nu$ with $\mu=6$ and $v_y=0$.]{
            \includegraphics[width=0.3\linewidth]{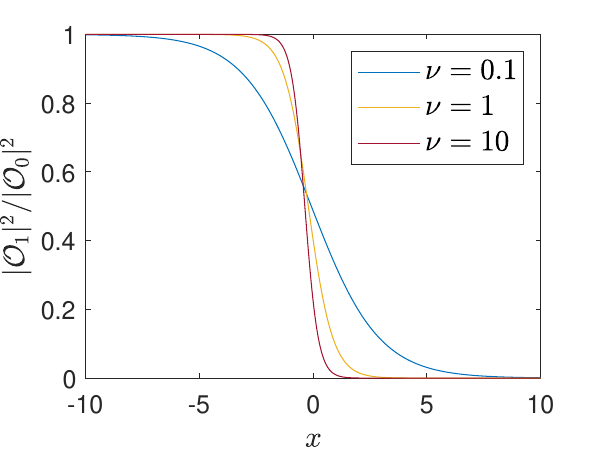}}
        \subfigure[Interface width $\delta$ and $|\mathcal{O}_0|^2$ for different $v_y$ with $\mu=6$ and $\nu=1$.]{
            \includegraphics[width=0.3\linewidth]{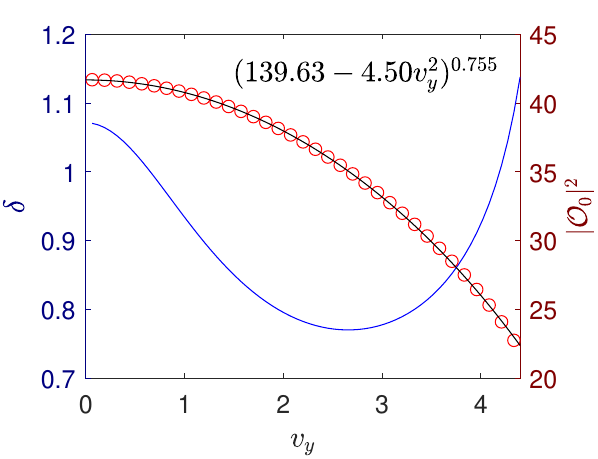}}
            \caption{Stationary configuration for immiscible binary superfluid. The normalized order parameter of the first superfluid component $\mathcal{O}_1$ for different \textbf{(a)} chemical potential $\mu$ and \textbf{(b)} coupling $\nu$, with $\mathcal{O}_0$ the value of the order parameter
            far from the interface. \textbf{(c)} The interface width $\delta$ (blue line) and $|\mathcal{O}_0|^2$ (red circles) for different relative velocity $v_y$.  Profiles of $|\mathcal{O}_2|^2$ are mirror image of those of $|\mathcal{O}_1|^2$ about $x=0$. Black line in \textbf{(c)} is the fitting result.}
    \label{profile}
\end{figure*}

%%%%%%%%%%%%%%%%%%%%%%%%%%%%%%%%%%%%%%%
\begin{figure*}[htpb]
        \centering
            \includegraphics[width=0.98\linewidth]{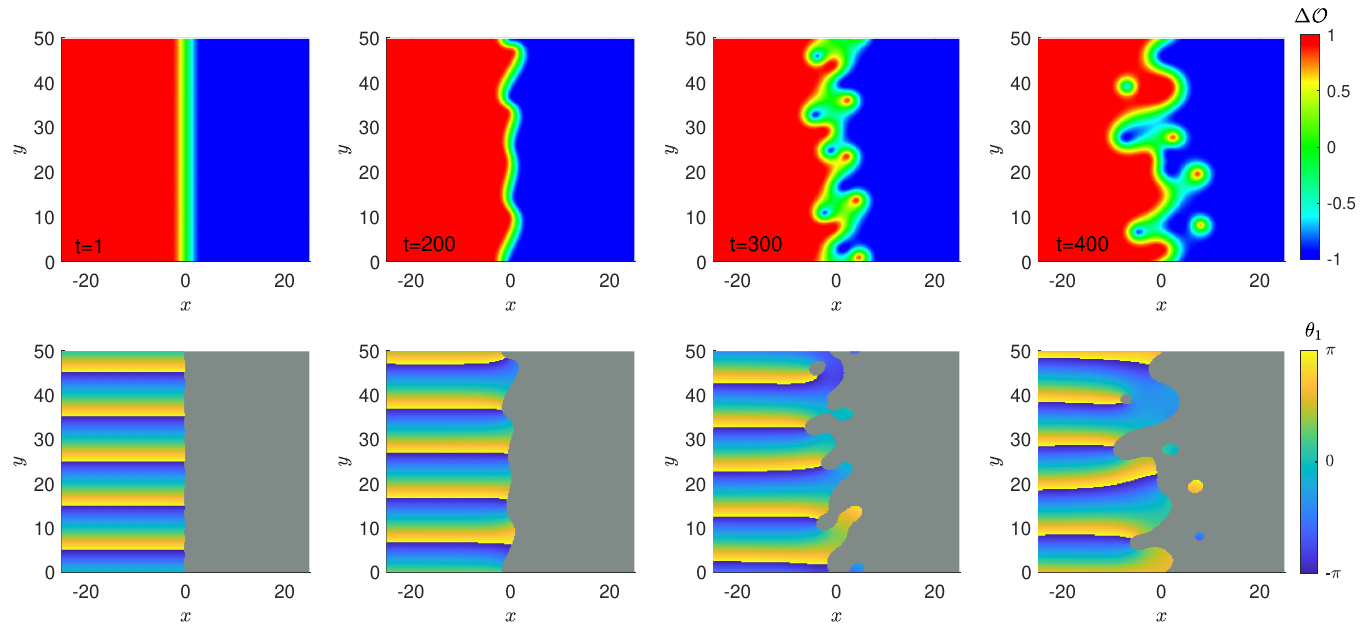}
            \caption{Interface dynamics for $v_y=1.2566$ at $T/T_c=0.677$. Snaps of condensation difference $\Delta\mathcal{O}=(|\mathcal{O}_1|^2-|\mathcal{O}_2|^2)/|\mathcal{O}_0|^2$ \textbf{(upper panel)} and the the profile of the phase of the first component $\theta_1$ \textbf{(bottom panel)} for different time are presented. Small initial perturbations on the interface destabilize and grow into larger amplitude structures leading to vortex formation and quantum turbulence. $\theta_1$ is only plotted for $|\mathcal{O}_1|^2-|\mathcal{O}_2|^2>0$ since otherwise $|\mathcal{O}_1|$ is small and $\theta_1$ would be pure noise. We choose $\nu=1$ and fix $\mu=6$.}
    \label{fancy}
\end{figure*}

\textbf{Stationary configuration--}
We begin with the stationary state describing an immiscible binary superfluid. Without loss of generality, we suppose that the two superfluid components undergo phase separation and form an straight
interface at $x=0$ together with a relative velocity $v_y$ along the $y$ axis. The corresponding bulk configuration are given by
\begin{equation}
\label{ansatz} \Psi_i=z\phi_i(z,x)e^{i\Theta_i(z,x,y)}, \quad A_t=A_t(z,x)\,,
\end{equation}
together with the gauge choice
\begin{equation}
\label{gauge}
    \partial_z\Theta_i=-\frac{A_t}{f}, \quad V_{iy}(z,x)\equiv\partial_y\Theta_i-A_y\,.
\end{equation}
We have $V_{ix}\equiv\partial_x\Theta_i-A_x=0$ for the stationary state. The phase $\theta_i$ of the condensation $\mathcal{O}_i$ is given by $\Theta_i|_{z=0}$ of~\eqref{ansatz}. Note that the phases $\theta_i$ depend on the $y$ coordinate because we will consider a relative velocity between the two superfluid components along the $y$-axis. According to the holographic dictionary, one has $V_{1y}|_{z=0}=-V_{2y}|_{z=0}=v_y/2$, where $v_y$ is the relative velocity between the two components. This results in a system of equations of motion involving 5 PDE’s for $(\phi_i, A_t, V_{iy})$ that all depend on the variables $z$ and $x$. We employ the Newton-Raphson method to solve the system. In $z$-direction, we use the Chebyshev pseudo-spectral method and, in $x$-direction, we adopt the fourth-order finite difference scheme and the Neumann boundary condition.  

The normalized profiles of $|\mathcal{O}_1|^2$ for different $\mu$ and $\nu$ with $v_y=0$ are shown in Fig.~\ref{profile} (a) and (b). Profiles of $|\mathcal{O}_2|^2$ are mirror image of those of $|\mathcal{O}_1|^2$ about $x=0$. We see generally that the larger $\mu$ and $\nu$ are, the narrower the interface is. This feature is qualitatively similar to the results from GPE (see~\cite{Yang:2019ibe} for an early study on the effect of $\nu$ in holographic superfluid with $\bm{v}=0$). The profiles can be fitted by
\begin{equation}
\label{fitformula}
    |\mathcal{O}_1|^2=\frac{|\mathcal{O}_0|^2}{2}(1-\mathrm{tanh}(x/\delta))\,,
\end{equation}
with $\delta$ the width of the interface and $\mathcal{O}_0$ the value of the condensation far from the interface. In GPE, when the coupling strength $\Delta$ between the two components is small, $\delta$ is given by $\delta=\xi/\Delta^{1/2}$\cite{QKH1}, where $\xi=\hbar/\sqrt{2m\mu}$ is the healing length. We have verified numerically that $\delta\sim(\mu-\mu_c)^{-1/2}$ and $\delta\sim\nu^{-1/2}$ when the coupling strength $\nu$ is small, which is reminiscent of the result from GPE. In constract, 
when the relative velocity is turned on, different behaviors from those of GPE appear, although the shape of condensate can still be fitted by~\eqref{fitformula}. As visible from Fig.~\ref{profile} (c), $\delta$ first decreases and then increases with $v_y$. Such behavior does not show in GPE. Besides, $|\mathcal{O}_0|^2$ versus $v_y$ can be fitted by $|\mathcal{O}_0|^2=(139.63-4.50 v_y^2)^{0.755}$, which also significantly deviates from the quadratic speed dependence from GPE.
 Nevertheless, the value of the power depends on the temperature, but is insensitive to the coupling $\nu$.

\textbf{Dynamical interface instability--}
We now provide the numerical simulations of the real-time dynamics. To initiate the dynamical instability, we give a small random noise to the initial stationary condensates. The system size $(L_x, L_y)$ is prepared properly for each parameter set such that the influence
of numerical boundaries can be omitted. Please refer to Appendix~\ref{app:s1} for numerical details.
\begin{figure*}[htpb]
        \centering
         \vspace{-0.35cm}
        \subfigtopskip=2pt
        \subfigbottomskip=2pt
        \subfigcapskip=-5pt
        \subfigure{
            \includegraphics[width=0.49\linewidth]{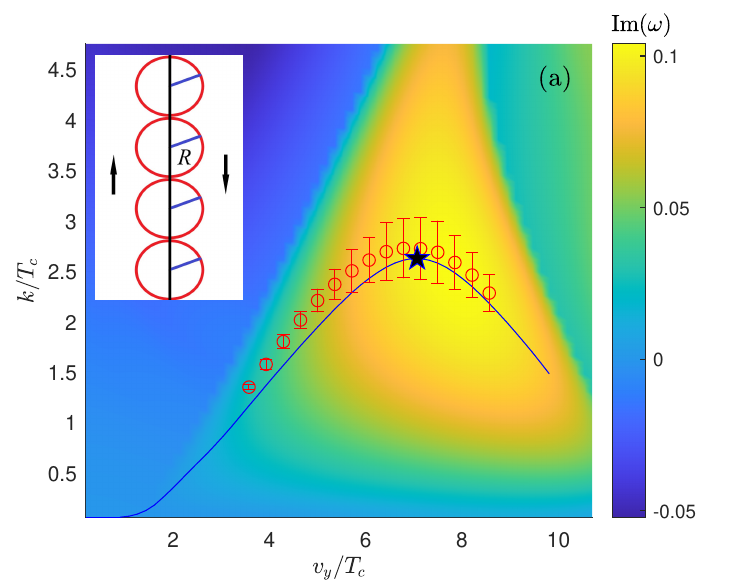}}
        \subfigure{
            \includegraphics[width=0.49\linewidth]{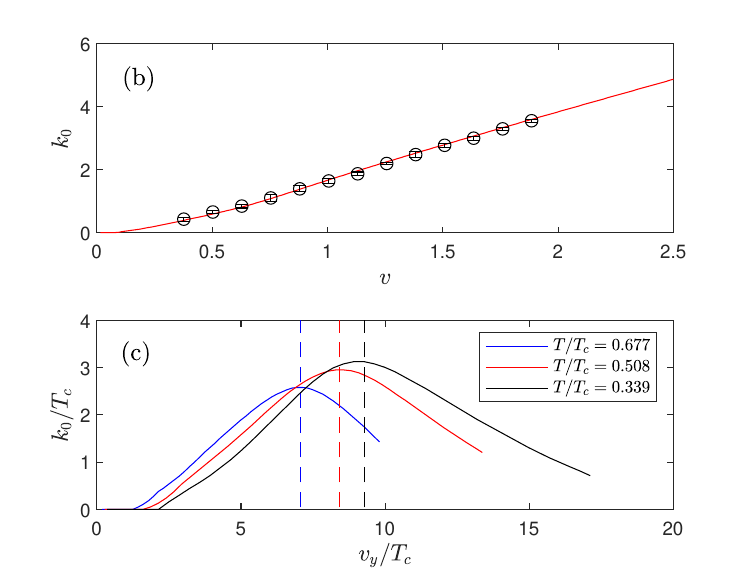}}
            \caption{\textbf{(a)} The wave number of the fastest growing mode $k_0$ versus the interfacial relative velocity for $T/T_c=0.677$. The circles with errorbar denote $k_0$ extracted from real time evolution and the solid line from perturbation analysis around the stationary state. The density plot gives the dominant QNMs for each wave number and velocity. \textbf{Insert} illustrates the highest point (star) at which the average distance of vortices $4\pi/v_y$ is equal to the vortex diameter $2R$. \textbf{(b)} $k_0-v$ relation of GPE at zero temperature. \textbf{(c)} The relation from our holographic theory~\eqref{action} at  different temperatures. Dash vertical lines show the critical velocities by equating the average distance of vortices generated along the interface and the vortex size. We choose $\nu=1$.}
    \label{k-v}
\end{figure*}

A representative example for the interface dynamics is shown in Fig.~\ref{fancy}. Far-from-equilibrium dynamics that are reminiscent of quantum KHI are manifest, just like what happens to the two-component superfluid interface under perturbations from solving GPE~\cite{QKH2,crossover,pattern}. The amplitude of the sinusoidal interface wave is monotonically
increased due to the exponential growth of initial perturbations. The
subsequent nonlinear evolution exhibits a quite different behavior
from that of the classical fluid. In particular, the interface eventually disintegrates
into bubble-like domains of the condensates. As visible from the phase profile $\theta_1$ of the first component, each bubble-like domain contains a quantized vortex (see the bottom panel at $t=400$). Similar dynamics can be found for different temperatures and relative velocities.

To gain the quantitative feature of the system, we consider the wave number of the most unstable modes $k_0$ versus the relative velocity $v_y$. This can be measured directly from experiments.  To obtain the mode $k_0$ that blows up most rapidly, we extract Fourier spectrum of the shape of interface at each time before the vortices develop. For each $v_y$, we find a stable peak in Fourier spectrum at $k_0$ during non-linear evolution. In practice, for each velocity we run 60 simulations with different perturbations and compute the mean value of $k_0$. The results are denoted by the red circles with errorbar in Fig.~\ref{k-v}(a). Surprisingly, one finds that there is a turning point (denoted as star) for the $k_0-v_y$ curve. 

The onset of the instability can be uncovered from linear response analysis around the stationary state with an interface layer (see Fig.~\ref{profile}) by calculating the quasi-normal modes (QNMs), see Appendix~\ref{app:QNM} for more details. Since the stationary solution possesses the time translation symmetry and translation symmetry along $y$ direction, one decomposes small perturbations in terms of Fourier modes $e^{-i(\omega t-ky)}$ where $\omega$ and {\color{red} $k$} represent the frequency and wave number of the interface wave, respectively. The quasi-normal frequencies generically take a complex value due to dissipations into the normal component. Once the imaginary part Im($\omega)>0$, the background becomes dynamically unstable. The wave number of the fastest growing mode corresponds a positive Im($\omega$) that takes a maximum at $k=k_0$. As shown by the density plot of Im($\omega$) in Fig.~\ref{k-v}(a), for a given $v_y$, the imaginary part rises with the increase of $k$, peaks at a certain wave number that corresponds to the fastest growing mode (see also Fig.~\ref{fig:omega} in Appendix~\ref{app:QNM}). The velocity dependence of the wave number from such dominant QNMs is denoted by the solid blue curve in Fig.~\ref{k-v}(a). One can see that the linear analysis agrees quantitatively with the one extracted from fully dynamical evolution. The slight deviation may be due to the relative wide instability spectrum and late-time nonlinear effects.

As a comparing, we do the same analysis by solving GPE (see Appendix~\ref{app:C} for more details). The results are shown in Fig.~\ref{k-v}(b), from which $k_0\sim v^2$ for small $v$ and $k_0\sim v$ for large $v$. This behavior is in sharp contrast to our holographic results of Fig.~\ref{k-v}(a). Another difference is that holographic simulation yields quite small $k_0$ in the small velocity regime. A possible reason might be that no dissipation and temperature effect are considered in GEP.

A heuristic picture for the non-monotonic behavior in Fig.~\ref{k-v}(a) is given as follows. Thanks to the quantum nature of superfluids, in particular the vortex quantization, the number of vortex formation along the interface is approximately given by $N=\frac{\Delta\theta}{2\pi}=\frac{v_y L_y}{4\pi}$ for our present system. Therefore, the average distance of vortices along the interface is estimated statistically as $L_0=L_y/N=4\pi/v_y$. Meanwhile, it is anticipated to create more vortices for a large wave number of the interface instability~\cite{vortex}. 
On the other hand, the vortex size can be obtained from the static vortex configuration which is axisymmetric and the condensation depends only on the radial coordinate (see \emph{e.g.}~\cite{Keranen:2009re}). As $v_y$ is increased, the average distance of vortices $L_0$ will decrease. At a critical velocity $v_c=2\pi/R$ for which $L_0=2R$ with $R$ the radius of a vortex, the vortices near the interface become so dense that they immediate contact with each other, see Insert of Fig.~\ref{k-v}(a) for illustration. The nonlinear vortex dynamics becomes important and prevent the increase of more vortices from the interface instability. Therefore, the corresponding value of the fastest growing mode $k_0$ at $v_c$ is the maximal wave number among all unstable modes. The above heuristic analysis agrees quantitatively with our numerical computations. Defining the radius of a single vortex $R$ at which $|\mathcal{O}(R)|^2/|\mathcal{O}_0|^2=0.98$, we get $v_c/T_c=2\pi/(RT_c)=7.06$ for $T/T_c=0.677$, matching exactly with the turning point in Fig.~\ref{k-v}(a). This is also confirmed for other temperatures, see Fig.~\ref{k-v}(c). Notice that the value of $v_y$ associated with the turning point does not correspond to the one for $\delta$ depicted \emph{e.g.} in Fig.~\ref{profile} (c). We highlight that the velocity range shown in Fig.~\ref{k-v} is below the critical velocity given by the Landau criterion and thus our interface dynamics is not due to the Landau instability~\cite{Amado:2013aea,2020arXiv201006232L,Gouteraux:2022qix,2023arXiv231208243A}.To have a better understanding of the interface instabilities, it will be helpful to work out the thermodynamics of a binary superfluid. Exploring the thermodynamics of inhomogenous binary superfluids and the thermodynamic instabilities is challenging and lies beyond the scope of our present investigation. 

%This is the main prediction of our holographic two-component superfluid model, which might be accounted for by strong coupling limit rooted in holographic duality.

\textbf{Discussion--}In this work, we study the interface dynamics of two-component superfluids at strong coupling. The interface separating the two phases of superfluid becomes unstable as the relative velocity is increased. The pattern observed from fully nonlinear simulation is reminiscent of quantum KHI. From both the far-from-equilibrium evolution and the linear QNMs analysis, we find that the wave number of the most unstable modes depends non-monotonically on the superfluid velocity, in sharp contrast to the results of GPE. We have uncovered that the turning point occurs when the mean separation of vortices generated by interface instabilities becomes comparable to the size of vortices, suggesting that the non-monotonicity is due to the direct interaction between neighbor vortices.
Moreover, as visible from Fig.~\ref{k-v}(a)(c), the instability develops noticeable only above a threshold value, which might be due to the dissipation and viscous effect away from the ground state. These findings are directly testable in platform, like strongly coupled ultracold Bose gases or thin helium films at low temperatures.

%Such predictions show great promise to be tested in the real experiments.

Our study broadens the application of holography to non-equilibrium phenomena with finite temperature and dissipation. In particular, it initiates the investigation of interface instabilities in the holography laboratory, providing an intriguing platform to explore the interplay of instabilities and the emergence of complex flow phenomena. There are several lines of research in which our study can be extended, shedding light on the complicated behaviors of interface dynamics. For example, due to the relative velocity between superfluid component and normal component, one anticipates that interface of binary superfluids that move with the same velocity relative to normal component can be unstable at finite temperature~\cite{QKH3}. Since GPE is at zero temperature, this phenomenon can not be presented from GPE. In contrast, the temperature effect is naturally incorporated in our holographic model, and, indeed, this is the case and will be reported elsewhere. Moreover, quantized vortices with higher winding number can develop~\cite{Lan:2023gyc}, which could complicate the interface dynamics. The turbulent dynamical behavior is anticipated following closely the initial emission of vortex-antivortex pairs. Moreover, the introduction of external magnetic field and rotation is of interest. \\

\begin{acknowledgements}
    We thank Fanlong Meng, Tao Shi, Hongbao Zhang and Yu Tian for useful comments and suggestions. This work is partly supported by the National Natural Science Foundation of China No.12122513, No.12075298 and No. 12275233. We acknowledge the use of the High Performance Cluster at Institute of Theoretical Physics, Chinese Academy of Science.
\end{acknowledgements}

%\printbibliography{}
% \newpage
% \onecolumngrid
\appendix 
% \clearpage
% \renewcommand\thefigure{S\arabic{figure}}    
% \setcounter{figure}{0} 
% \renewcommand{\theequation}{S\arabic{equation}}
% \setcounter{equation}{0}
% \renewcommand{\thesubsection}{SI\arabic{subsection}}
\section*{Appendices}

Here we provide additional technical details on the derivation of the main results reported in the main text.

\section{Numerical scheme of the fully non-linear simulations}\label{app:s1}

The general equations of motion for our matter fields read
 \begin{equation}
    \mathcal{D}_\mu \mathcal{D}^\mu\Psi_i-m_i^2\Psi_i-\frac{\nu}{2}|\Psi_j|^2\Psi_i=0, \quad(i,j=1,2\quad i\ne j),
 \end{equation}
 \begin{equation}
     \nabla_\mu F^{\mu\nu}=-2\mathrm{Im}(\sum_i\Psi^*_i\mathcal{D}^\nu\Psi_i).
 \end{equation}
where $\mathrm{Im}$ represents imaginary part.
The bulk equations of motion on top of the background~\eqref{backg} are explicitly given by
\begin{widetext}

    \begin{equation}
        \begin{aligned}
            \label{phi}
            2\partial_t\partial_z\Phi_i-[2i A_t\partial_z\Phi_i+i \partial_zA_t\Phi_i+\partial_z(f\partial_z\Phi_i)-z\Phi_i
            +\partial_x^2\Phi_i+\partial_y^2\Phi_i
            -i (\partial_xA_x+\partial_yA_y)\Phi_i&\\
            -(A_x^2+A_y^2)\Phi_i-2i (A_x\partial_x\Phi_i+A_y\partial_y\Phi_i)
            -\frac{\nu}{2}|\Phi_j|^2\Phi_i]=0, \qquad(i,j=1,2\quad i\ne j)&
        \end{aligned}
    \end{equation}
    \begin{equation}
        \label{At}
        \begin{aligned}
            \partial_t\partial_zA_t-[\partial_x^2A_t+\partial_y^2A_t+f\partial_z(\partial_xA_x+\partial_yA_y)-\partial_t(\partial_xA_x+\partial_yA_y)
            -2A_t\sum_i|\Phi_i|^2&\\
            -2f\mathrm{Im}(\sum_i\Phi_i^*\partial_z\Phi_i)+2\mathrm{Im}(\sum_i\Phi_i^*\partial_t\Phi_i)]=0\,,&
        \end{aligned}
    \end{equation}
    \begin{equation}
        \label{Ax}
        \begin{aligned}
            2\partial_t\partial_zA_x-[\partial_z(\partial_xA_t+f\partial_zA_x)+\partial_y(\partial_yA_x-\partial_xA_y)-2A_x\sum_i|\Phi_i|^2
            +2\mathrm{Im}(\sum_i\Phi_i^*\partial_x\Phi_i)]=0\,,
        \end{aligned}
    \end{equation}
    \begin{equation}
        \label{Ay}
        \begin{aligned}           
        2\partial_t\partial_zA_y-[\partial_z(\partial_yA_t+f\partial_zA_y)+\partial_x(\partial_xA_y-\partial_yA_x)-2A_y\sum_i|\Phi_i|^2
            +2\mathrm{Im}(\sum_i\Phi_i^*\partial_y\Phi_i)]=0\,,
        \end{aligned}
    \end{equation}
        \begin{equation}
        \label{constraint}
        \begin{aligned}     
        \partial_z(\partial_xA_x+\partial_yA_y-\partial_zA_t)-2\mathrm{Im}(\sum_i\Phi_i^*\partial_z\Phi_i)=0\,,
        \end{aligned}
    \end{equation}
            
\end{widetext}
where $\Phi_i=\Psi_i/z$. For simplicity, we have chosen $m_1^2=m_2^2=-2/L^3$, $e_1=e_2=L=1$ and have adopted the radial gauge $A_z=0$. Notice the last equation is a constraint with no time derivative. These equations are not independent. They obey the following constraint equation
    \begin{equation}
    \label{relation}
    \begin{aligned}
        -\partial_t\mathrm{Eq.}(\ref{constraint})-\partial_z\mathrm{Eq.}(\ref{At})+\partial_x\mathrm{Eq.}(\ref{Ax})+\partial_y\mathrm{Eq.}(\ref{Ay})\\=2\mathrm{Im}(\sum_i\mathrm{Eq.}(\ref{phi})\times\Phi_{0i}^*)\,.     
    \end{aligned}
    \end{equation}

The expansions of the fields near the AdS boundary $z=0$ can be obtained as
    \begin{equation}
            A_\mu=a_\mu+b_\mu z+\cdots,\quad \Phi_i=(\Phi_i)_s+(\Phi_i)_v z+\cdots.
    \end{equation}
From the holographic duality, the coefficients $a_t$, $a_i$ $(i = x,y)$ and $(\Phi_i)_0$ are interpreted as the chemical potential $\mu$, vector potential, and scalar operator source of the boundary theory, respectively. To describe superfluid where the U(1) symmetry is broken spontaneously, we should turn off the scalar source, \emph{i.e.} $(\Phi_i)_s=0$. Then the superfluid condensation is given as $\mathcal{O}_i=(\Psi_i)_v$ in the standard quantization. 

Note that $T$ and $\mu$ are not independent quantities because of scaling symmetry of the system. After fixing $z_h=1$, $\mu$ is the only free parameter. Then, there is a second order phase transition for our present setup when $\mu\ge \mu_c\simeq 4.064$. This also fixes the ratio $T/T_c=\mu_c/\mu$. In practice, we fix $T=3/4\pi$, so $T_c=\mu T/\mu_c$ should depend on $\mu$.

The fully non-linear simulation starts with the initial data
\begin{equation}
 \Phi={\Phi}_{i0}+\delta\Phi_i,\quad A_\mu={A}_{\mu 0}+\delta A_\mu\,,
\end{equation}
where ${\Phi}_{i0}$ and ${A}_{\mu 0}$ denote the corresponding profile for the stationary interface configuration (see \emph{e.g.} Fig.1 in the main text). For simplicity but without loss of generality, we use a sum of evenly distributed modes as initial condition:
\begin{equation}
\begin{aligned}
        \delta\Phi_i&=(-1)^{i-1}\sum_k z\mathrm{exp}(-x^2)\mathrm{exp}(iky+i\theta_k)\mathrm{exp}(i(v_i)_y),\\ \delta A_\mu&=0\,,
\end{aligned}
\end{equation}
where $\theta_k$ is a random phase for each wave number $k$.

Our evolution scheme is implemented numerically by the fourth order Runge-Kutta method along the time direction.  Moreover, we use Chebyshev pseudo spectral method along the $z$ direction and 
Fourier pseudo spectral method along the $y$ coordinate. To capture the dynamics near the interface, we adopt the fourth order finite difference scheme in the $x$ direction. Previous holographic investigations deal exclusively with periodic boundary condition along $x$ direction and therefore cannot properly accommodate the interface dynamics.

First, we use \eqref{phi}, \eqref{Ax} and \eqref{Ay} to evolve $\Phi$, $A_x$ and $A_y$ subject to the source free boundary condition at the AdS boundary:
\begin{equation}
\Phi(z=0)=A_x(z=0)=A_y(z=0)=0\,.    
\end{equation}
together with the Neumann boundary condition 
\begin{equation}
\partial_x\Phi=\partial_x A_\mu=0\,,
\end{equation}
at $x=\pm L_x/2$, where the system size $L_x$ is prepared properly for each parameter set such that the influence of numerical boundaries can be omitted. Note that period boundary condition has been implicitly adopted in our numerical computation.

Then we use \eqref{At} to evolve $\partial_zA_t$ on the AdS boundary. Since we set the chemical potential $\mu$ as a constant, $-\partial_zA_t(z=0)$ is just the charge density $\rho$ of the dual boundary system. Finally, $A_t$ can be solved by the constraint equation~\eqref{constraint} subject to the boundary condition
\begin{equation}
\partial_zA_t(z=0)=-\rho, \quad A_t(z=0)=\mu\,.
\end{equation}
The later time configuration can be obtained in the same way as described before.

\section{Linear instability around a stationary configuration}\label{app:QNM}

The onset of the instability of such stationary solutions can be analyzed by the linear response theory. To be more specific, we turn to the linear perturbations on the stationary background.
\begin{equation}
\begin{aligned}
     \Phi_i=\Phi_{i0}+\delta\Phi_i,\quad A_t=A_{t0}+\delta A_t, \\ \quad A_x=A_{x0}+\delta A_x, \quad A_y=A_{y0}+\delta A_y\,, 
\end{aligned}
\end{equation}
where $\Phi_{0i}$, $A_{t0}$, $A_{x0}$ and $A_{y0}$ are stationary solutions as shown in Fig.\,1 in the main text.
Taking into account the translation invariance of our  background along the time and $y$ directions, as well as the velocity difference across the interface between two superfluids, one takes the bulk perturbation fields as the form of
\begin{widetext}

    \begin{equation}
    \begin{aligned}
        &\delta\Phi_i=u_i(z,x)e^{-i(\omega t-ky)}e^{i(v_i)_yy}, \quad \delta\Phi_i^*=v_i(z,x)e^{-i(\omega t-ky)}e^{-i(v_i)_yy},\\&\delta A_t=a_t(z,x)e^{-i(\omega t-ky)},\quad\delta A_x=a_x(z,x)e^{-i(\omega t-ky)}, \quad\delta A_y=a_y(z,x)e^{-i(\omega t-ky)} ,
    \end{aligned}
    \end{equation}
        
\end{widetext}
where $(v_1)_y=-(v_2)_y=v_y/2$ with $v_y$ the relative velocity between the two superfluid components. The resulting linear perturbation equations are given explicitly as
\begin{widetext}

   \begin{equation}
    \begin{aligned}
        \label{pertubation}        &2iA_{t0}\partial_zu_i+2ia_t\partial_z\Phi_{0i}+i\partial_zA_{t0}u_i+i\partial_za_t\Phi_{0i}
        +\partial_z(f\partial_zu_i)-zu_i +\partial_x^2u_i-(k+(v_i)_y)^2u_i
        -i\partial_xA_{x0}u_i\\
        &-i\Phi_{0i}(\partial_xa_x+ika_y)-(A_{x0}^2+A_{y0}^2)u_i-2A_{x0}\Phi_{0i}a_x-2A_{y0}\Phi_{0i}a_y
        -2i(A_{x0}\partial_xu_i+i(k+(v_i)_y)A_{y0}u_i)\\&-2i(a_x\partial_x\Phi_{0i}+ia_y(v_i)_y\Phi_{0i})-\frac{\nu}{2}|\Phi_{0j}|^2u_i
        -\frac{\nu}{2} \Phi_{0j}^*\Phi_{0i}u_j-\frac{\nu}{2}  \Phi_{0j}\Phi_{0i}v_j\\&=-2i\omega\partial_zu_i, \qquad(i,j=1,2\quad i\ne j)\,,\\
        &-2iA_{t0}\partial_zv_i-2ia_t\partial_z\Phi_{0i}^*-i\partial_zA_{t0}v_i-i\partial_za_t\Phi_{0i}^*
        +\partial_z(f\partial_zv_i)-zv_i +\partial_x^2v_i-(k-(v_i)_y)^2v_i
        +i\partial_xA_{x0}v_i\\
        &+i\Phi_{0i}^*(\partial_xa_x+ika_y)-(A_{x0}^2+A_{y0}^2)v_i-2A_{x0}\Phi_{0i}^*a_x-2A_{y0}\Phi_{0i}^*a_y
        +2i(A_{x0}\partial_xv_i+i(k-(v_i)_y)A_{y0}v_i)\\&
        +2i(a_x\partial_x\Phi_{0i}^*-ia_y(v_i)_y\Phi_{0i}^*)-\frac{\nu}{2}|\Phi_{0j}|^2v_i
        -\frac{\nu}{2} \Phi_{0j}\Phi_{0i}^*v_j-\frac{\nu}{2}  \Phi_{0j}^*\Phi_{0i}^*u_j\\&=-2i\omega\partial_zv_i, \qquad(i,j=1,2\quad i\ne j)\,,\\
        &\partial_x^2a_t-k^2a_t+f\partial_z\partial_xa_x+ikf\partial_za_y-2a_t\sum_i|\Phi_{0i}|^2-2A_{t0}\sum_i(\Phi_{0i}^*u_i+\Phi_{0i}v_i)+if\sum_i(\Phi_{0i}^*\partial_zu_i\\
        &-\Phi_{0i}\partial_zv_i+v_i\partial_z\Phi_{0i}-u_i\partial_z\Phi_{0i}^*)=-i\omega(\partial_za_t+\partial_xa_x+ika_y)+\omega\sum_i(\Phi_{0i}^* u_i-\Phi_{0i} v_i)\,,\\
        &\partial_z(\partial_xa_t+f\partial_za_x)-(k^2a_x+ik\partial_xa_y)-2a_x\sum_i|\Phi_{0i}|^2-2A_{x0}\sum_i(\Phi_{0i}^*u_i+\Phi_{0i}v_i)\\
        &-i\sum_i(\Phi_{0i}^*\partial_xu_i-\Phi_{0i}\partial_xv_i+v_i\partial_x\Phi_{0i}-u_i\partial_x\Phi_{0i}^*)=-2i\omega\partial_za_x,\\
        &ik\partial_za_t+\partial_z(f\partial_za_y)+\partial_x^2a_y-ik\partial_xa_x-2a_y\sum_i|\Phi_{0i}|^2-2A_{y0}\sum_i(\Phi_{0i}^*u_i+\Phi_{0i}v_i)\\
        &+\sum_i((k+(v_i)_y)\Phi_{0i}^*u_i-(k-(v_i)_y)\Phi_{0i}v_i+(v_i)_yv_i\Phi_{0i}+(v_i)_yu_i\Phi_{0i}^*)=-2i\omega\partial_za_y\,.
    \end{aligned}
   \end{equation} 
       
\end{widetext}

For more stable numerical performance, we use the following equation for $a_t$:
\begin{equation}
    \begin{aligned}
    &\partial_z(\partial_xa_x+ika_y-\partial_za_t)+i\sum_i(\Phi_i^*\partial_zu_i+v_i\partial_z\Phi_i\\&-u_i\partial_z\Phi_i^*-\Phi_i\partial_zv_i)=0,
    \end{aligned}
\end{equation}
which comes from the constraint equation~\eqref{constraint}. Moreover, we require the last perturbation equation of~\eqref{pertubation} to be satisfied at the AdS boundary $z=0$, yielding
\begin{equation}
(\partial_z\partial_xa_x+ik\partial_za_y=-i\omega \partial_za_z)|_{z=0}\,. 
\end{equation}
Then, by considering~\eqref{relation}, the last perturbation equation is also satisfied in the whole bulk.  Regarding other perturbed fields, we impose the source free boundary condition at the AdS boundary. We further consider the Neumann boundary condition at $x=\pm L_x/2$. 

 \begin{figure}[htpb]
        \centering
            \includegraphics[width=0.99\linewidth]{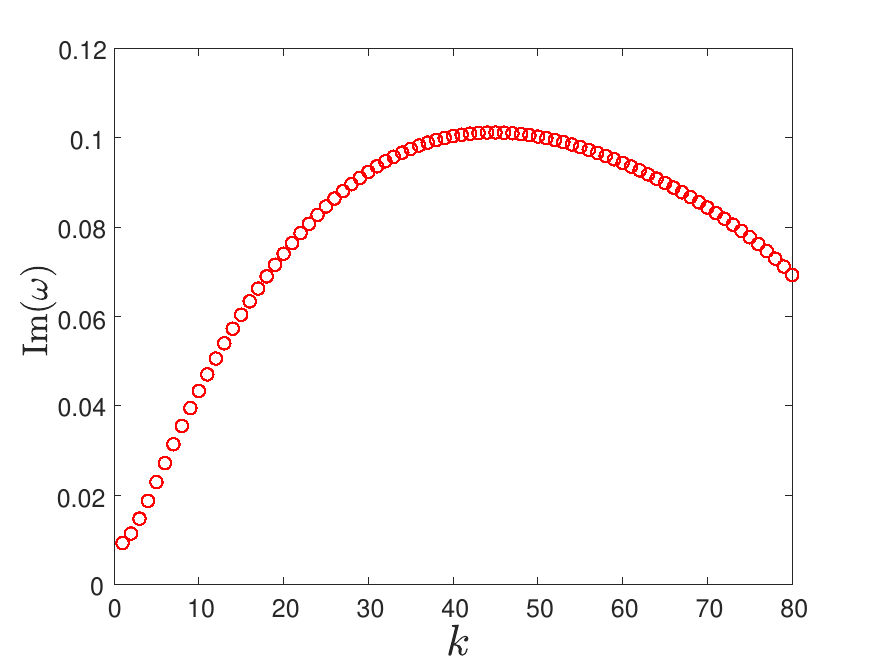}
            \caption{The imaginary of the low lying spectrum of QNMs of stationary configurations with $T/T_c=0.677$ and $v_y=2.5132$. The peak of Im($\omega$) determines the wave number of the fastest growing mode. the We fix $\nu=1$ and $\mu=6$.}
    \label{fig:omega}
\end{figure}

The corresponding quasi-normal modes are extracted by solving the above generalized eigenvalue problem. Then we can  numerically obtain $\omega$ for each $k$ and velocity difference $v_y$. Due to dissipations into the normal component, the quasi-normal frequencies generically take a complex value. Since $\delta\Phi_i\sim e^{-i\omega t}$, the stationary configuration will become dynamically unstable whenever $\mathrm{Im}(\omega)>0$. The larger the positive imaginary part is, the more unstable the system becomes. In Fig.~\ref{fig:omega}, we demonstrate the spectrum of QMNs versus $k$ for $v_y=3.5132$ at $T/T_c=0.677$ and $\nu=1$. The imaginary part rises with the increase of $k$, peaks at a certain wave number that corresponds to the fastest growing mode. The case for the linear perturbation analysis of GPE is exactly the same, but much simpler. 

\section{Interface instability from GPE}
\label{app:C}
Two-component superfluids can be described by following coupled GPEs~\cite{PhysRev.134.A543}:
    \begin{equation}
    \label{GP_2}
    \begin{aligned}
        i\partial_t \Psi_i=(-\frac{1}{2m_i}\nabla^2-\mu_i+g_i|\Psi_i|^2+g_{ij}|\Psi_j|^2+V_i)\Psi_i,\\ \quad (i,j=1,2, \quad i\ne j)\,.      
    \end{aligned}
    \end{equation}
    Here we focus on $g_{12}>\sqrt{g_1g_2}$, which gives immiscible BECs, and we also set $V_i=0$. To study two-component superfluids with interface and 
    relative velocity, we use the ansatz
    \begin{equation}
    \label{ansatz_GP}
        \Psi_i(\mathbf{r})=\psi_i(x)e^{im_iv_iy}\,,
    \end{equation}
with $v_1=-v_2=v$. Substitute (\ref{ansatz_GP}) into (\ref{GP_2}), we get the following time-independent GPE for $\psi_i(x)$
\begin{equation}
\begin{aligned}
    (-\frac{1}{2m_i}\partial_x^2-\mu_i+\frac{m_iv^2}{2}+g_i|\psi_i|^2+g_{ij}|\psi_j|^2)\psi_i=0\,, \\ \quad (i,j=1,2, \quad i\ne j)\,.    
\end{aligned}
\end{equation}
Far from the interface, we have $\partial_y\psi_i=0$, $\psi_j=0$ and therefore $\psi_i=(\mu_i-m_iv^2/2)/g_i=\sqrt{n_i}$. Profiles for $\psi_i$ solved from
these equations are similar to those shown in Fig.~\ref{profile}. 

   \begin{figure}[htpb]
        \centering
            \includegraphics[width=1\linewidth]{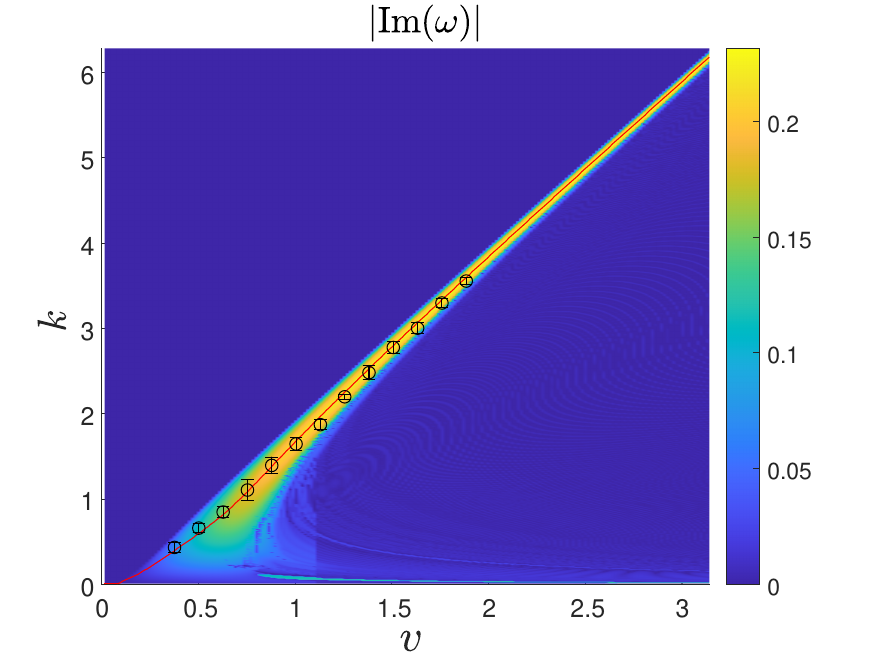}
            \caption{The wave number of the fastest growing mode versus the superflow velocity $v_1=-v_2=v$ obtained from GPEs. The red solid line is from perturbation analysis of~\eqref{BdG} and the black circles with error bars are extracted from dynamical evolution. The density plot shows the dominant $|\text{Im}(\omega)|$ for each wave number and velocity. We have chosen $g=m=1$, $g_{12}=2$ and $\mu=0.5$.}
    \label{v_k_GP}
\end{figure}

Next let's study the interface instability using linear perturbation analysis. Adding perturbations on the 
stationary background $\psi_i^0$
\begin{equation}
    \Psi_i=\left[\psi_i^0(x)+u_i(x)e^{iky-i\omega t}-w_i^*(x)e^{-iky+i\omega^* t}\right]e^{im_iv_iy}\,,
\end{equation}
and linearizing GPEs, we obtain the BdG equation:
\begin{equation}\label{BdG}
    \mathcal{H}\mathbf{U}=\omega\mathbf{U}\,,
\end{equation}
\begin{equation}
    \mathcal{H}=\left\{
\begin{tabular}{cccc}
$h_1^+$&$-g_1(\psi_1^0)^2$&$g_{12}\psi_1^0\psi_2^0$&$-g_{12}\psi_1^0\psi_2^0$\\
$g_1(\psi_1^0)^2$&$-h_1^-$&$g_{12}\psi_1^0\psi_2^0$&$-g_{12}\psi_1^0\psi_2^0$\\
$g_{12}\psi_1^0\psi_2^0$&$-g_{12}\psi_1^0\psi_2^0$&$h_2^+$&$-g_2(\psi_2^0)^2$\\
$g_{12}\psi_1^0\psi_2^0$&$-g_{12}\psi_1^0\psi_2^0$&$g_2(\psi_2^0)^2$&$-h_2^-$
\end{tabular}
\right\},
\end{equation}
with $\mathbf{U}=(u_1,w_1,u_2,w_2)^\mathrm{T}$ and 
\begin{equation}
    h_i^\pm=-\frac{1}{2m_i}\left[\partial_x^2-(k\pm m_iv_i)^2\right]-\mu_i+2g_i|\psi_i^0|^2+g_{ij}|\psi_j^0|^2.
\end{equation}
By numerically diagonalizing this discretized BdG Hamiltonian $\mathcal{H}$, we can get the eigenfrequency $\omega$. Since this Hamiltonian $\mathcal{H}$ is real, we also have $\mathcal{H}\mathbf{U}^*=\omega^*\mathbf{U}^*$, \emph{i.e.}, $\omega^*$ is also an eigenvalue whenever $\omega$ is an eigenvalue. Therefore,
whenever $\mathrm{Im}(\omega)\ne 0$, 
the system is dynamically unstable. For convenience, we set $g_{12}=2$, $g_1=g_2=1$, $m_1=m_2=1$, and $\mu_1-m_1v_1^2/2=\mu_2-m_2v_2^2/2=\mu=0.5$ when 
doing numerical calculations.

By calculating $\mathrm{Im}(\omega)$ for different $v$ and $k$, we can extract the wave number $k_0$ of the most unstable mode for each $v$,
which can be compared with the results from our holographic model. We have also done the fully nonlinear time evolution to extract $k_0$ with the same procedure as in the holographic case. Results are shown in Fig.~\ref{v_k_GP}. We see the values of $k_0$ extracted from both linear analysis and dynamical evolution agree well with each other.
From Fig.~\ref{v_k_GP}, we also find that $k_0\sim v^2$ for small $v$ and $k_0\sim v$ for large $v$, corresponding to the Kelvin-Helmholtz instability and the counter-superflow instability, respectively, as shown in~\cite{crossover}. This result is qualitatively different from the one from our holographic model (see Fig.~\ref{k-v}).

\bibliography{refs}
\end{document}